\documentclass[12pt]{article}

\usepackage[margin=1in]{geometry}
\usepackage{setspace}
\usepackage{booktabs}
\usepackage{graphicx}
\usepackage{multirow}
\usepackage{longtable}
\usepackage{lscape}
\usepackage{algorithm}
\usepackage{algpseudocode}
\usepackage{natbib}
\usepackage{hyperref}
\usepackage{titling}
\usepackage{amsmath,amsthm,amssymb,amsfonts,amsbsy,bm,latexsym}
\usepackage{dsfont}
\usepackage{mathrsfs}
\usepackage{xcolor}

\hypersetup{colorlinks=true,linkcolor=blue,citecolor=blue,urlcolor=blue}
\graphicspath{{Fig/}}

\def\a{{\bf a}}
\def\b{{\bf b}}
\def\A{{\bf A}}
\def\B{{\bf B}}
\def\C{{\bf C}}

\def\x{{\bf x}}

\def\z{{\bf z}}

\def\I{{\bf I}}
\def\P{{\bf P}}

\def\vv{{\bf v}}

\def\mR{\mathbb{R}}

\def\mE{E}
\def\scrE{\mathscr{E}}

\def\calA{\mathcal{A}}

\def\calM{\mathcal{M}}

\def\calT{\mathcal{T}}

\def\calL{\mathcal{L}}

\def\mR{\mathbb{R}}

\def\mE{E}
\def\scrE{\mathscr{E}}

\def\calA{\mathcal{A}}

\def\calM{\mathcal{M}}

\def\calT{\mathcal{T}}

\def\calL{\mathcal{L}}

\def\bbeta{{\bm \beta}}

\def\bome{{\bm \omega}}

\newcommand{\trans}{^{\mbox{\tiny{T}}}}

\def\wh{\widehat}
\def\wt{\widetilde}
\newcommand{\norm}[1]{\Vert#1\Vert}

\newcommand{\inner}[2]{\langle #1, #2 \rangle}
\newcommand{\Inner}[2]{\left\langle #1, #2 \right\rangle}
\newcommand{\abs}[1]{\vert#1\vert}

\newcommand{\bSig}{\bm{\Sigma}}

\def\cov{\operatorname{cov}}

\def\supp{\operatorname{supp}}
\def\DHBIC{\operatorname{DHBIC}}
\DeclareMathOperator*{\argmin}{arg\,min}

\newtheorem{theorem}{\bf Theorem}
\newtheorem{lemm}{\bf Lemma}

\newtheorem{remark}{\bf Remark}
\newtheorem{assum}{\bf Assumption}

\newcommand{\calN}{\mathcal{N}}
\newcommand{\bzeros}{\bm{0}}

\def\ora{\mbox{\tiny{ora}}}
\def\ARE{\operatorname{ARE}}

\title{\bfseries Sparse Rank Regression for Restricted-Access Economic Data}
\author{
Wen Zhang, Songshan Yang, Liping Zhu\thanks{Corresponding author: zhu.liping@ruc.edu.cn}\\[0.3em]
\small Center for Applied Statistics and Institute of Statistics and Big Data,\\ \small Renmin University of China\\
}
\date{}

\begin{document}
\onehalfspacing
\maketitle

\begin{abstract}
Empirical research in economics increasingly relies on restricted-access data held by multiple firms or agencies, making it impossible to construct the estimator of interest on the pooled sample. At the same time, heavy-tailed distributions are pervasive in economics and finance outcomes such as prices, expenditures and loan sizes. 
We study sparse, robust estimation in the restricted-access setting. The infeasible pooled benchmark is convoluted rank regression (CRR), a smooth rank-based estimator
designed for heavy-tailed outcomes. 
Because the CRR criterion is a non-additive $U$-statistic, existing communication-efficient methods built for additive empirical losses do not directly apply. We propose distributed convoluted rank regression (DCRR), a surrogate criterion built from a single local CRR loss and an aggregated gradient correction, and show that it shares the same population minimizer as the pooled CRR objective.
Building on this surrogate, we develop a two-stage sparse procedure: an iterative $\ell_1$-penalized stage followed by a folded-concave
refinement. For the resulting estimator, we establish non-asymptotic error bounds, a distributed strong oracle property, and a distributed
criterion for consistent model selection.
Simulations and an application to used-car prices show that DCRR closely approximates pooled CRR and improves on naive divide-and-conquer, particularly under heavy-tailed errors.
\end{abstract}

\noindent\textbf{Keywords:} communication efficiency; administrative and proprietary data; heavy-tailed outcomes; rank regression; restricted-access data

\medskip
\noindent\textbf{JEL classification:} C13; C14; C21; C55; C83

\bigskip

\section{Introduction}


Administrative, financial, platform, and proprietary datasets are the basis of empirical work in economics, yet disclosure rules, competitive concerns, and privacy constraints routinely prevent the central pooling of individual records \citep{einav2014economics,meyer2015surveys,abowd2019privacy}. 
The econometric challenge is to estimate
a single pricing, forecasting, or policy relation from data that are physically and legally distributed across firms or agencies. This paper studies sparse, robust estimation of such a relation in a restricted-access environment. 
In this setting, only summary statistics, not raw records, may cross institutional boundaries.

A motivating application is the estimation of a common hedonic price equation from dealership-level used-car transactions. Hedonic vehicle-price equations date back to \citet{griliches1961hedonic} and \citet{rosen1974hedonic}, and used-car markets are a canonical setting in which quality heterogeneity and information frictions matter for pricing \citep{akerlof1970lemons,genesove1993adverse}.
Consider a lender financing vehicle purchases, an insurer pricing guaranteed-value contracts, or an online marketplace producing valuation tools. At the level of any such institution, individual transaction records remain held by the originating dealer, and contractual, competitive, or privacy constraints can make record-level pooling infeasible. The econometric target is nevertheless a common sparse pricing equation that can be used for collateral valuation, underwriting, residual-value forecasting, or quality-adjusted price measurement.
This setting also illustrates why robustness matters. Even within a relatively homogeneous vehicle segment, transaction prices reflect unobserved vehicle condition, local demand, trade-in terms, and inventory pressure, so the conditional price distribution is often asymmetric and may exhibit long tails. Section~\ref{section:5} returns to this dealer-level application. 

As the application above illustrates, restricted access is only one part of the problem; two additional features of these data make estimation challenging. First, modern economic datasets often contain many controls, categorical indicators, interactions, and flexible transformations, so that the number of potential predictors can be large relative to the sample size and sparse estimation becomes essential \citep{fan2011sparse,belloni2014inference}. 
Second, tail behavior is economically important: heavy-tailed distributions are pervasive in economics and finance \citep{gabaix2009power}, and empirical outcomes such as prices, expenditures, revenues, loan sizes, and earnings are often skewed and may contain occasional extreme observations. In such settings, procedures based on conditional means can be sensitive to a relatively small number of extreme values, whereas rank-based and quantile methods provide more robust descriptions of heterogeneous economic relations \citep{jurevckova1971nonparametric,jaeckel1972estimating,koenker1978regression,koenker2005quantile,hettmansperger2010robust}. 

Motivated by these considerations, we study a sparse linear location model with exogenous regressors and a common coefficient vector across data sites. The object of interest is the coefficient vector that would be obtained from the pooled robust estimator if unrestricted access to the full sample were available. For heavy-tailed and asymmetric outcomes, rank-based regression is a natural candidate for this target because it is insensitive to monotone transformations and outliers in the response \citep{jurevckova1971nonparametric,jaeckel1972estimating,hettmansperger2010robust}.
Building on this tradition, \citet{zhou2024sparse} proposed convoluted rank regression (CRR), a smooth rank-type $U$-statistic loss that retains the robustness of traditional rank estimators while enjoying differentiability and favorable efficiency, and established oracle properties for its penalized version. Related sparse rank regression methods are studied in \citet{fan2020rank} and \citet{mei2024lasso}; for sparse robust $M$-estimation, see \citet{loh2017statistical} and \citet{wang2020tuning}.

The main difficulty under restricted access is that the CRR objective is a second-order $U$-statistic: the pooled criterion contains cross-site pairs that are not available to any single data holder, so it cannot be written as an average of local losses. Most communication-efficient procedures, by contrast, are built for additive empirical criteria, whether through averaging local estimators \citep{zhang2013divide,rosenblatt2016optimality,battey2018distributed} or iterative gradient exchange \citep{jordan2019communication,wang2017efficient,fan2023communication}; see also \citet{chen2014split,lee2017communication,huang2019distributed}. Existing distributed procedures for robust and quantile regression have likewise been developed for additive loss structures \citep{chen2020distributed,Pan02102022,shi2020communication}. 
However, there are currently no theoretical guarantees for estimators under non-additive U-statistic losses, where the global empirical loss cannot be decomposed as an average of local losses. 
Filling this gap is one of the main goals of the present
paper.

To overcome this non-additive challenge in restricted-access settings, we propose distributed convoluted rank regression (DCRR), based on a surrogate criterion formed by a single local CRR loss and an aggregated gradient correction. The surrogate is not equal to the pooled CRR criterion, but it has the same population minimizer. This population-level alignment is sufficient to obtain first-order equivalence to pooled CRR under suitable conditions. Building on this surrogate, we develop a two-stage DCRR procedure. In the first stage, we compute an $\ell_1$-penalized DCRR estimator via an iterative distributed algorithm that communicates only gradients of local CRR losses. In the second stage, we refine this initial estimator by applying a folded-concave penalty through a local linear approximation scheme, in the spirit of nonconvex penalization \citep{fan2001variable,zhang2010nearly} and the I-LAMM framework \citep{Fan2015ILAMMFS}. 
We further develop a distributed HBIC criterion for tuning parameter selection and show that it consistently recovers the true support of $\bbeta^\ast$ in a distributed environment.


The paper contributes in three respects. First, on the methodology side, it extends communication-efficient estimation to a non-additive rank criterion by exploiting population-level alignment between the local surrogate and the pooled objective; to our knowledge, this is the first treatment of distributed second-order $U$-statistic losses in a sparse, high-dimensional setting. Second, on the theory side, it develops a two-stage regularized DCRR procedure and establishes non-asymptotic error bounds together with a distributed strong oracle property; with $M$ data sites, total sample size $N$, sparsity $s$, and ambient dimension $p$, the theory permits $M=o\{N/(s^{2}\log p)\}$ while requiring only $O(\log N)$ rounds of communication. Third, on the implementation side, it introduces a distributed HBIC criterion and proves its model-selection consistency, so that the procedure is fully data-driven. For empirical work, the results show when a researcher can recover essentially the same sparse robust relation as pooled CRR without access to the pooled microdata.

\medskip
The rest of the paper is organized as follows. Section~\ref{section:2} introduces the model, the DCRR surrogate loss, and the two-stage DCRR iterative algorithm. Section~\ref{section:3} establishes non-asymptotic error bounds, the distributed strong oracle property, and the consistency of the distributed HBIC selector. Section~\ref{section:4} reports Monte Carlo evidence, and Section~\ref{section:5} presents an empirical application to used-car prices. Section~\ref{sec:discussion} concludes.

\subsection{Notation}

We introduce some necessary notation. 
For $a,b\in\mR$, $a\wedge b=\min\{a,b\}$ and $a\vee b=\max\{a,b\}$. For any positive semi-definite matrix $\A$, $\lambda_{\min}(\A)$ and $\lambda_{\max}(\A)$ represent the smallest and the largest eigenvalues.
For a vector $\a$, $\supp(\a)=\{j\mid a_j\neq 0\}$. For a positive integer $p$, $[p]=\{1,2,\ldots,p\}$.
For an index set $\calA$, a vector $\b$ and matrix $\C$, $\b_{\calA}=(b_j,j\in\calA)$ and $\C_{\calA\calA}=(c_{ij},i\in\calA ,j\in\calA)$. 
For a vector $\a=(a_1,\ldots,a_p)$ and $q\in[1,\infty)$, define the $\ell_q$ norm as $\norm{\a}_q=(\sum_{j=1}^p |a_j|^q)^{\frac{1}{q}}$. Let $\norm{\a}_{\infty}=\max_{j}\abs{a_j}$ be the $\ell_{\infty}$ norm and $\norm{\a}_{\min}=\min_{j}\abs{a_j}$.
For a symmetric positive-definite matrix $\B$, define the vector norm as $\norm{\a}_{\B}=\norm{\B^{1/2}\a}_2$.
For a sequence $a_n$ and another nonnegative sequence $\{b_n\}$,
we write $a_n=O(b_n)$ when there exists a constant $C>0$ such that $\abs{a_n}\leq Cb_n$; we use $a_n=o(b_n)$ when $a_n/b_n\rightarrow0$. We write $a_n\lesssim b_n$ when there exists a constant $C>0$ such that $a_n\leq Cb_n$; we write $a_n\asymp b_n$ if $a_n\lesssim b_n$ and $b_n\lesssim a_n$. 
For two random sequences $Z_n$ and $Z_n'$, we write $Z_n=O_p(Z_n')$ when for any $\epsilon>0$, there exists a constant $M>0$ and $N$, $\P(\abs{Z_n/Z_n'}>M)<\epsilon$ for all $n\geq N$; we write $Z_n=o_p(Z_n')$ when $\lim_{n\rightarrow\infty}\P(\abs{Z_n/Z_n'}>\epsilon)=0$, for any $\epsilon>0$.

\section{Econometric setup and estimator}
\label{section:2}
In this section, we formalize the restricted-access setting and introduce the distributed convoluted rank regression (DCRR) estimator. In applications such as dealer-level transaction data, each site retains its own records and only derived summaries can be transmitted to a master site. 
We first review the infeasible pooled convoluted rank regression (CRR) criterion and then construct the surrogate loss and sparse estimators used in the sequel.

\subsection{Model and convoluted rank regression}

We consider a restricted-access setting where outcomes such as transaction prices, loan amounts, or revenues are stored on $M$ sites. For notational convenience, we present our methodology and theory under a balanced design in which each site holds the same number $n$ of observations so that the total sample size satisfies $N = nM$; 
The balanced design is not necessary, and we discuss the details of heterogeneous local sample sizes in Remark~\ref {rem:unequal-n}. 
Suppose we observe identically distributed data $\{(y_{i}^{m},\x_{i}^{m}): i=1,\ldots,n,\; m=1,\ldots,M\}$,
where $\{(y_i^{m},\x_i^{m})\}_{i=1}^n$ denotes the subsample stored on the $m$-th site. Here $y_{i}^m\in \mR$ is the response and $\x_{i}^m\in\mR^p$ is the $p$-dimensional covariate vector; in the motivating application, they correspond to a transaction price and observed vehicle characteristics. We write the pooled sample as $\{(y_i,\x_i)\}_{i=1}^N$. 

Assume that the data are generated from the linear regression model
\begin{equation}
\label{equation:1}
y_i^m=\x_i^m\bbeta^\ast+\epsilon_i^m,
\quad i=1,\ldots,n,\; m=1,\ldots,M,
\end{equation}
where $\{\epsilon_i^m: i=1,\ldots, n, m=1,\ldots,M\}$ are i.i.d.\ random errors, and $\bbeta^\ast\in \mR^p$ is the unknown parameter vector. Without loss of generality, we assume that the errors in \eqref{equation:1} have mean zero and are independent of $\x$. 
All our results can extend to the more general case with unequal local sample sizes. 

Our analysis is conducted under a sparse linear model with exogenous regressors and i.i.d.\ sampling across observations and sites. Accordingly, the paper is concerned with estimation and model selection for a common conditional location relation, rather than with endogeneity, sample selection, cluster dependence, or fixed effects. These maintained conditions are standard in extremum-estimation analysis \citep{newey1994large} and delimit the scope of the results.

Let $(y',\x')$ be an independent copy of $(y,\x)$. For rank regression, the true regression coefficient $\bbeta^\ast$ can be characterized as the minimizer of the population rank loss
\begin{equation}
  \label{eq:min_1}
\bbeta^\ast=\argmin_{\bbeta\in\mR^p}\mE\big\{\abs{y-y'-(\x-\x')\trans\bbeta}\big\}.
\end{equation}
However, the non-smooth absolute loss brings substantial challenges for computation and theoretical  analysis. To address this, \citet{zhou2024sparse} proposed the canonical convoluted rank regression. The loss function is defined as:
\begin{equation}
\label{equation:2}
\min_{\bbeta\in \mR^p} \frac{1}{N(N-1)}\sum_{i=1}^{N}\sum_{j\neq i}
L_h\big\{y_i-y_j-(\x_i-\x_j)\trans\bbeta\big\},
\end{equation}
where
$L_h(u)=\int_{-\infty}^{\infty}\abs{u-v}\frac{1}{h}K\!\left(\frac{v}{h}\right)\,dv$
is a smooth convex function obtained as the convolution of $L(u)=\abs{u}$ and $K_h(u)=\frac{1}{h}K(u/h)$. 

We assume that the kernel $K$ satisfies the following properties:  
(i) $K(-t)=K(t)$ for all $t \in \mathbb{R}$;  
(ii) there exists $\delta_0>0$ such that $\kappa_l:=\inf _{t \in[-\delta_0, \delta_0]} K(t)>0$;  
(iii) $\int_{-\infty}^{\infty} K(t) \mathrm{d} t=1$;  
(iv) $\kappa_u:=\sup _{t \in \mathbb{R}} K(t)<\infty$;  
(v) $\kappa_j:=\int_{-\infty}^{\infty}|t|^j K(t) \mathrm{d} t<\infty$ for $j=1,2$;  
(vi) there exist $\alpha_0 \in(0,1]$ and $L_0>0$ such that 
$|K(x)-K(y)| \leq L_0|x-y|^{\alpha_0}$ for all $x, y \in \mathbb{R}$.  
Typical examples include the Gaussian kernel $K(u)=\frac{1}{\sqrt{2\pi}}\exp\{-u^2/2\}$ and the Epanechnikov kernel $K(u)=\frac{3}{4}(1-u^2)\I(-1\leq u \leq 1)$.

Let $\mathcal L_h(\bbeta)=\mE\big\{L_h\big(y-y'-(\x-\x')\trans\bbeta\big)\big\}$ and denote its minimizer by $\bbeta_h^\ast$. \citet{zhou2024sparse} showed that for any $h>0$, $\bbeta_h^\ast=\bbeta^\ast$,
so the smoothing does not introduce bias at the population level.

\subsection{Distributed CRR and surrogate loss}

In a restricted-access system, the full CRR loss in \eqref{equation:2} is not directly available, because it involves all $N(N-1)$ pairs across sites. We define the global and local empirical losses as
\begin{align*}
\calL_N(\bbeta)
&=\frac{1}{N(N-1)}\sum_{i=1}^{N}\sum_{j\neq i}
L_h\big(y_i-y_j-(\x_i-\x_j)\trans\bbeta\big),\\
\calL_m(\bbeta)
&=\frac{1}{n(n-1)}\sum_{i=1}^{n}\sum_{j\neq i}
L_h\big(y_i^m-y_j^m-(\x_i^m-\x_j^m)\trans\bbeta\big),
\quad m=1,\ldots,M.
\end{align*}
The global loss $\calL_N$ is infeasible to compute in a restricted-access environment when $N$ is large, whereas each $\calL_m$ is computable on the $m$-th local site.

For general $M$-estimators with an additive empirical loss of the form
\[
\calL_N(\bbeta)=\frac{1}{M}\sum_{m=1}^M \calL_m(\bbeta),
\]
\citet{jordan2019communication} proposed the communication-efficient surrogate likelihood (CSL) framework. In its simplest form, CSL constructs the surrogate loss
\begin{equation}
\label{equation:3}
\wt{\calL}(\bbeta)=\calL_1(\bbeta)
-\Inner{\bbeta}{\nabla\calL_1(\bbeta_0)-\nabla\calL_N(\bbeta_0)},
\end{equation}
where $\bbeta_0$ is an initial estimator for $\bbeta^\ast$ and the first site is used as the center. This construction relies on the additivity $\nabla\calL_N(\bbeta)=M^{-1}\sum_{m=1}^M\nabla\calL_m(\bbeta)$, which allows the global gradient to be reconstructed from local gradients. Under additive losses, sample-level additivity is therefore a convenient route to a valid surrogate because it automatically guarantees that the surrogate and the full-data objective are aligned both empirically and at the population level.

In convoluted rank regression, however, the empirical loss is a second-order $U$-statistic over all sample pairs. The global CRR loss $\calL_N$ couples observations across sites through cross-site pairs, whereas each $\calL_m$ only involves within-site pairs. As a consequence,
\[
\frac{1}{M}\sum_{m=1}^M\nabla\calL_m(\bbeta)\neq \nabla \calL_N(\bbeta),
\]
so the additivity assumption underlying CSL fails. We therefore construct a surrogate loss directly for CRR. Our analysis relies on the surrogate and the pooled CRR criterion sharing the same population minimizer, even though their sample representations differ.

Inspired by the form of \eqref{equation:2} and the CSL idea in \eqref{equation:3}, we propose the distributed convoluted rank regression (DCRR) and the loss function is
\[
\calL_d(\bbeta;\bbeta_0)
:=\calL_1(\bbeta)
-\inner{\bbeta}{\nabla\calL_1(\bbeta_0)-\frac{1}{M}\sum_{m=1}^M\nabla\calL_m(\bbeta_0)}.
\]
which uses the CRR loss on the master site together with a gradient correction aggregated from the local sites. The distributed convoluted rank regression estimator is defined as:
\begin{equation}
  \label{equation:dcr}
  \wh\bbeta_d=\argmin_{\bbeta\in\mR^p}
  \Big\{
  \calL_1(\bbeta)
  -\inner{\bbeta}{\nabla\calL_1(\bbeta_0)-\frac{1}{M}\sum_{m=1}^M\nabla\calL_m(\bbeta_0)}
  \Big\},
  \end{equation}
By construction, $\calL_d(\cdot;\bbeta_0)$ is not equal to the infeasible full-data loss $\calL_N$, and $\frac{1}{M}\sum_{m=1}^M\nabla\calL_m(\bbeta)$ does not coincide with $\nabla\calL_N(\bbeta)$. The next result shows, however, that $\calL_d$ and $\calL_N$ have the same population minimizer. This property underlies the asymptotic equivalence between DCRR and centralized CRR.

Let \begin{equation}\label{eq:min_2}\bbeta_d^\ast=\argmin_{\bbeta\in\mR^p}\mE\{\calL_d(\bbeta;\bbeta_0)\}.\end{equation} The following lemma shows that the population minimizer of our surrogate loss coincides with the true parameter $\bbeta^\ast$.

\begin{lemm}
\label{lemm:1}
For any $h>0$ and $M\geq 1$, the minimizers $\bbeta^\ast$ and $\bbeta_d^\ast$ defined in \eqref{eq:min_1} and \eqref{eq:min_2}, respectively, satisfy $\bbeta_d^\ast=\bbeta^\ast$.
\end{lemm}

This conclusion follows directly from the equality $\bbeta_h^\ast = \bbeta^\ast$ for the population CRR loss, and it provides a population-level justification for our surrogate construction.

We next consider the asymptotic distribution of the DCRR estimator. The following lemma shows that, under a mild condition on the initial estimator, $\wh\bbeta_d$ is asymptotically equivalent to the centralized CRR estimator.

\begin{lemm}
\label{lemm:2}
For each $h>0$, if $\norm{\bbeta_0-\bbeta^\ast}_2=o_p(M^{-1/2})$, then
\[
\sqrt{N}(\wh\bbeta_d-\bbeta^\ast)
\xrightarrow{d}\calN\left(
\bzeros,\,
\frac{\mE\big[\mE\{L_h'(\epsilon-\epsilon')\mid\epsilon\}^2\big]}{\mE\{L_h''(\epsilon-\epsilon')\}^2}
\cov(\x)^{-1}
\right),
\]
where $(\epsilon,\epsilon')$ is an independent copy of the error pair and $\cov(\x)=\bSig$ is the covariance matrix of $\x$.
\end{lemm}

\begin{remark}
Lemma~\ref{lemm:2} shows that, under the initial condition $\norm{\bbeta_0-\bbeta^\ast}_2=o_p(M^{-1/2})$, the DCRR estimator $\wh\bbeta_d$ has the same asymptotic variance as the global convoluted estimator $\wh\bbeta_h=\argmin_{\bbeta}\calL_N(\bbeta)$. Let $\wh\bbeta=\argmin_{\bbeta}\frac{1}{N(N-1)}\sum_{i=1}^N\sum_{j\neq i}\abs{y_i-y_j-(\x_i-\x_j)\trans\bbeta}$ be the nonsmoothed rank estimator. Then the asymptotic relative efficiency satisfies
\[
\ARE(\wh\bbeta_d,\wh\bbeta)
=\frac{\mE\big[\mE\{L_h'(\epsilon-\epsilon')\mid\epsilon\}^2\big]}{12\left\{\int f^2(x)\,dx\right\}^2\mE\{L_h''(\epsilon-\epsilon')\}^2},
\]
where $f(\cdot)$ is the density of $\epsilon$.
From Theorem~3 of \citet{zhou2024sparse}, $\ARE(\wh\bbeta_d,\wh\bbeta)\to 1$ as $h\to 0^+$;
thus, choosing $h$ sufficiently small makes the efficiency of $\wh\bbeta_d$
arbitrarily close to that of $\wh\bbeta$.
Together with the theoretical results in Section~\ref{section:3}, this shows that
DCRR achieves centralized CRR efficiency while using only local $U$-statistics
and $O(\log N)$ rounds of communication.
\end{remark}

\subsection{Sparse DCRR and iterative algorithms}
In many economic applications, the dimension $p$ can grow with $n$ as empirical specifications include many controls, categorical indicators, interactions, and flexible transformations, while only a relatively small subset of coordinates of $\bbeta^\ast$ has non-negligible effects on the outcome.
Let $\calA=\{j: \beta_j^\ast\neq 0\}$ be the support of $\bbeta^\ast$ and $s=\abs{\calA}\geq1$ be its size. We allow both $p$ and $s$ to diverge with $n$, and assume that $s$ is of smaller order than $n$.

To incorporate sparsity, we consider the penalized DCRR estimator by solving
\begin{equation}
\label{pen-dcrr}
\min_{\bbeta\in \mR^p} \left\{
\calL_d(\bbeta;\bbeta_0)
+\sum_{j=1}^{p}p_{\lambda}(\abs{\beta_j})
\right\},
\end{equation}
where $p_{\lambda}(\cdot)$ is a penalty function with tuning parameter $\lambda$, such as the $\ell_1$ penalty or a folded-concave penalty. 
We now describe a two-stage iterative algorithm that performs $\ell_1$-penalized iterations in the first stage, followed by a second-stage refinement based on folded-concave penalties. Theoretical properties of these estimators are presented in Section~\ref{section:3}.

In the first stage, we perform an initial iteration using the DCRR with an $\ell_1$ penalty.
We design the $\ell_1$-penalized DCRR iterative algorithm as
\begin{equation}
\label{eq:l1_stage}
{\wh{\bbeta}}^{k}:=\argmin_{\bbeta\in\mR^p}
\Big\{
\calL_1(\bbeta)
-\Inner{\bbeta}{\nabla\calL_1({\wh{\bbeta}}^{k-1})-\frac{1}{M}\sum_{m=1}^M\nabla\calL_m({\wh{\bbeta}}^{k-1})}
+\lambda\sum_{j=1}^{p}\abs{\beta_j}
\Big\},
\end{equation}
for $k=1,\ldots,k_1$, where $k_1\geq1$ is the maximum number of iterations. At iteration $k$, we use the previous estimate $\wh{\bbeta}^{k-1}$ as the initial value in the DCRR loss. The central site broadcasts $\wh{\bbeta}^{k-1}$ to all local sites, each local site computes the gradient $\nabla\calL_m(\wh{\bbeta}^{k-1})$, and the center aggregates these gradients to form the surrogate loss. Each iteration incurs a communication cost of order $O(pM)$, so the $\ell_1$-penalized DCRR algorithm is communication-efficient.

Motivated by the strong oracle properties of folded-concave penalized estimators \citep{fan2001variable,zhang2010nearly}, we employ a second-stage refinement based on folded-concave penalties to enhance estimation accuracy and achieve distributed strong oracle guarantees. 

For the folded-concave penalized DCRR, we adopt a local linear approximation (LLA) \citep{zou2008one} to $p_{\lambda}(\cdot)$ and consider penalties of the form $p_\lambda(v)=\lambda^2 p(v / \lambda)$ for $v \geq 0$, where $p:[0, \infty) \rightarrow[0, \infty)$ satisfies:  
(i) $p(\cdot)$ is non-decreasing and concave on $[0, \infty)$ with $p(0)=0$;  
(ii) $p^{\prime}(\cdot)$ exists almost everywhere, is non-increasing on $(0, \infty)$, $0 \leq p^{\prime}(v) \leq 1$, and $\lim _{v \downarrow 0} p^{\prime}(v)=1$;  
(iii) $p^{\prime}(\alpha_1)=0$ for some $\alpha_1>0$.  
Special cases include the SCAD penalty \citep{fan2001variable} and the MCP penalty \citep{zhang2010nearly}.

For convenience, we denote the $k$-th iterate of the $\ell_1$-penalized DCRR algorithm by $\wh\bbeta^{(1,k)}$. Let the initial estimator for the second stage be $\wh{\bbeta}^{(2,0)}=\wh{\bbeta}^{(1,k_1)}$, and let $k_2$ be the maximum number of iterations in the second stage. Using the LLA idea, for $t\geq 2$ and $k=1,\ldots,k_t$, we define the $t$-th stage estimator $\wh{\bbeta}^{(t,k)}$ as
\begin{equation}
\label{eq:multi_stage}
\wh{\bbeta}^{(t,k)}=\argmin_{\bbeta\in\mR^p}
\Big\{
\calL_d(\bbeta;\wh\bbeta^{(t,k-1)})
+\sum_{j=1}^{p} p^{\prime}_{\lambda_{t,k}}\big(\abs{\wh{\beta}_j^{(t,0)}}\big)\abs{\beta_{j}}
\Big\},
\end{equation}
where $\wh{\bbeta}^{(t,0)}=\wh\bbeta^{(t-1,k_t)}$. In practice, we will mainly focus on the case $k_t=1$ for $t\geq2$, which already suffices to achieve the distributed strong oracle property described in Section~\ref{section:3}. The corresponding simplified second-stage algorithm is summarized in Algorithm~\ref{algo:2}.

\begin{algorithm}[t] 
\caption{Two-Stage DCRR Iterative Algorithm}
\label{algo:2}
\begin{algorithmic}[1]

\State \textbf{Input:} bandwidth $h$, maximum number of iterations $k_1$, maximum number of stages $T$, initial value $\wh{\bbeta}^0=\argmin_{\bbeta\in \mR^p}\Big\{\calL_1(\bbeta)+\lambda\sum_{j=1}^{p}\abs{\beta_j}\Big\}$.

\For{$k=1,\ldots,k_1$}
  \State \textbf{Center:} transmit the current iterate $\wh{\bbeta}^{(1,k-1)}$ to all local sites;
  \State \textbf{Local:} each site $m$ computes the local gradient $\nabla \calL_m(\wh{\bbeta}^{(1,k-1)})$ and sends it to the master site;
  \State \textbf{Center:} update the estimator by \eqref{eq:l1_stage}.
\EndFor

\For{$t=2,\ldots,T$}
   \State \textbf{Center:} set $\wh\bbeta^{(t,0)}=\wh\bbeta^{(t-1)}$;
   \State \textbf{Center:} transmit the current iterate $\wh{\bbeta}^{(t,0)}$ to all local sites;
   \State \textbf{Local:} each site $m$ computes the local gradient $\nabla \calL_m(\wh{\bbeta}^{(t,0)})$ and sends it to the master site;
   \State \textbf{Center:} update the estimator by 
   \Statex\[
        {\wh{\bbeta}}^{(t)}:=\argmin_{\bbeta\in\mR^p}
        \Big\{
        \calL_1(\bbeta)
        -\Inner{\bbeta}{\nabla\calL_1({\wh{\bbeta}}^{(t,0)})-\frac{1}{M}\sum_{m=1}^M\nabla\calL_m({\wh{\bbeta}}^{(t,0)})}
        +\sum_{j=1}^{p}p_{\lambda}'\big(\abs{\wh\beta_j^{(t,0)}}\big)\abs{\beta_j}
        \Big\}.
        \]
\EndFor\\
\Return $\wh\bbeta^{(T)}$.
\end{algorithmic}
\end{algorithm}

\section{Theoretical properties of sparse DCRR}
\label{section:3}

In this section, we establish non-asymptotic error bounds and oracle properties for the proposed sparse DCRR estimators. We first introduce regularity assumptions and notation, 
then study the convergence of the two-stage DCRR iterative algorithm,
the distributed oracle property, and tuning parameter selection via distributed HBIC.

\subsection{Assumptions and notation}

We begin with regularity conditions on the error distribution and the design.

\begin{assum}[Error distribution]
\label{assum:1}
Let $g(\cdot)$ be the probability density function of $\zeta_{ij}=\epsilon_i-\epsilon_j$. We assume that $g$ is Lipschitz continuous: there exists a constant $L_{1}>0$ such that 
$|g(x)-g(y)| \leq L_{1}|x-y|$ for all $x, y \in \mathbb{R}$. 
This implies $\mu_{0}:=\sup _{t \in \mathbb{R}} g(t)<\infty$. Moreover, we assume that there exist positive constants $\delta_{1}, \mu_{1}$ such that 
$g(t) \geq \mu_{1}$ for all $t \in[-\delta_{1}, \delta_{1}]$.
\end{assum}

\begin{assum}[Design]
\label{assum:2}
The covariate vector $\x$ has bounded components and zero mean:
$\norm{\x}_{\infty} \leq b_1$ almost surely and $\mE(\x)=\bzeros$. Let $\x'$ be an independent copy of $\x$ and define $\z=\bSig^{-1/2}(\x-\x')$, where $\bSig=\mE(\x\x\trans)$. We assume that $\z$ is sub-Gaussian, i.e., there exists a constant $v_1\geq 1$ such that
$\P\big(\abs{\z\trans\z'}>v_1\norm{\z'}_2 t\big)\leq 2e^{-t^2/2}$
for all $\z'\in\mR^p$ and $t\geq 0$.
\end{assum}

\begin{assum}[Eigenvalue condition]
\label{assum:3}
There exists a constant $\rho>0$ such that 
$\min_{\vv\in\mR^p}\frac{\norm{\vv}_{\bSig}^2}{\norm{\vv}_2^2}\geq \rho$,
where $\norm{\vv}_{\bSig}^2=\vv\trans\bSig\vv$. In particular, $\lambda_{\min}(\bSig)\geq \rho$ and $\lambda_{\max}(\bSig)<\infty$.
\end{assum}
Assumption~\ref{assum:1} is a Lipschitz condition on the density of the error difference which only imposes local regularity on the density of $\epsilon_i-\epsilon_j$ and does not require moment conditions on the errors. Hence it is also compatible with heavy tail error, such as Student's $t$ and Cauchy type errors.
Assumption~\ref{assum:2} is standard for random design, and we implicitly work with centered covariates so that no intercept is needed. The sub-Gaussian condition is used to derive sparse concentration bounds. Assumption~\ref{assum:3} is a restricted eigenvalue type condition that allows us to control the $\ell_2$ norm via the $\bSig$ norm through $\norm{\vv}_2\leq \rho^{-1/2}\norm{\vv}_{\bSig}$. In fact, we only require this to hold on certain cone sets introduced below. 

For $d,l>0$, we define the local elliptical region and $\ell_1$ cone
\[
\Theta(d)=\{\vv\in \mR^p:\, \norm{\vv}_{\bSig}\leq d\},
\qquad
\Lambda(l)=\{\vv\in\mR^p:\, \norm{\vv}_1\leq l\norm{\vv}_{\bSig}\}.
\]

We also introduce shrinkage factors for the 
folded-concave penalized DCRR estimators 
\begin{equation}
\label{def:gamma}
\gamma_1 \asymp \sqrt{\frac{s(\log p+x)}{n}},\qquad \gamma \asymp \sqrt{\frac{s+x}{nh}},\qquad \lambda^\ast\asymp \sqrt{\frac{\log p+x}{N}}.
\end{equation}
$x>0$ is a tuning parameter appearing in concentration bounds. 
We also assume that there exists $\alpha_0>0$ such that
\[
\phi=\frac{\big[1+\{p'(\alpha_0)/2\}^2\big]^{1/2}}{\alpha_0\kappa\rho^{1/2}}\in(0,1),
\]
and define
\[
l=\Big\{2+\frac{2}{p'(\alpha_0)}\Big\}
\Big(\frac{c^2+1}{\rho}\Big)^{1/2}s^{1/2},
\]
where $c>0$ satisfies $p'(\alpha_0)(c^2+1)^{1/2}/2+1=\alpha_0\kappa\rho^{1/2}c$. These constants will be used to characterize the contraction behavior of the iterative algorithms.

\begin{remark}[Different local sample sizes]
\label{rem:unequal-n}

More generally, suppose the $m$-th site contains $n_m$ observations with $N = \sum_{m=1}^M n_m$, and let $\calL_m$ be the corresponding local CRR loss based on $n_m$ samples. If the master site is chosen to be site $m_\star$, with local sample size $n_\star = n_{m_\star}$, then the DCRR surrogate loss and algorithms are defined exactly as before, except that $\calL_1$ is replaced by $\calL_{m_\star}$. All arguments in Section~\ref{section:3} carry over with $n$ replaced by $n_\star$ in the conditions of Lemma~\ref{theo:2} and Theorems~\ref{theo3}. The relevant requirement is that $n_\star$ satisfies the same dimensional scaling as $n$, for example $s^2 \log p / n_\star = o(1)$ and $M = o\{N/(s^2 \log p)\}$. Since $\max_{1\le m\le M} n_m \ge N/M$, we can always select the site with the largest local sample size as the master, ensuring that these conditions are met whenever the global scaling assumptions hold. 
\end{remark}

\subsection{Convergence of the two-stage DCRR iterative algorithm}
\label{section:3.2}
We now study the non-asymptotic error bounds of the two-stage DCRR iterative algorithm. 
For simplicity, let 
\[
r_{t,k}=\norm{\wh\bbeta^{(t,k)}-\bbeta^\ast}_{\bSig},
\qquad
r_{t,0}=\norm{\wh\bbeta^{(t,0)}-\bbeta^\ast}_{\bSig},
\]
and define the index set
$\calT_{t,k}
=\calA\cup\big\{j\in[p]:\lambda_j^{(t,k)}<p'(\alpha_0)\lambda_{t,k}\big\}$,
where $\mathbf{\lambda}^{(t,k)}=(\lambda_1^{(t,k)},\ldots,\lambda_p^{(t,k)})\trans$ with $\lambda_j^{(t,k)}=p^{\prime}_{\lambda_{t,k}}(\abs{\wh{\beta}_j^{(t,0)}})$. We assume that $\abs{\calT_{t,k}}\leq (c^2+1)s$, which is a sparsity condition on the initial estimator and is naturally satisfied by the $\ell_1$-penalized estimator.
Let $\bome^\ast=\frac{1}{M}\sum_{m=1}^M\nabla\calL_m(\bbeta^{\ast})$ and denote its restriction to $\calA$ by $\bome^\ast_{\calA}$.

\begin{lemm}
\label{theo:2}
Suppose that Assumptions~\ref{assum:1}--\ref{assum:3} hold and that $\abs{\calT_{t,k}}\leq (c^2+1)s$.
Let $h\gtrsim\{s\log p/n+\sqrt{x/n}\}$ and $\lambda_{t,k}\asymp\gamma_{1} r_{t,k-1} +\lambda^\ast$. For any $t\geq 1$ and $1\leq k\leq k_t$, if $\wh\bbeta^{(t,k-1)}-\bbeta^\ast\in\Lambda(l)$ and $\supp(\wh\bbeta^{(t,k-1)})\lesssim s$, then $\wh{\bbeta}^{(t,k)}$ satisfies: $\wh\bbeta^{(t,k)}-\bbeta^\ast\in\Lambda(l)$, $\supp(\wh\bbeta^{(t,k)})\lesssim s$, and
    \begin{equation}
    \label{theo:2equ}
    r_{t,k}
    \leq \gamma r_{t,k-1}+\phi r_{t,0}
    +\kappa^{-1}\rho^{-1/2}
    \Big\{
    \big\|\bome^\ast_{\calA}\big\|_2
    +  \big\|p_{\lambda_{t,k}}^{\prime}\big((\abs{\bbeta_{\calA}^\ast}-\alpha_0\lambda_{t,k})_{+}\big)\big\|_2
    \Big\},
    \end{equation}
with probability at least $1-e^{-x}$.
\end{lemm}

\begin{remark}
From \eqref{theo:2equ}, we obtain
\[
r_{t,k} 
\leq \{\gamma^k+\phi(1-\gamma^k)/(1-\gamma)\}r_{t,0}
+\frac{1-\gamma^k}{1-\gamma}\,
\kappa^{-1}\rho^{-1/2}
\Big\{
\big\|\bome^\ast_{\calA}\big\|_2
+\big\|p_{\lambda_{t,k}}^{\prime}\big((\abs{\bbeta_{\calA}^\ast}-\alpha_0\lambda_{t,k})_{+}\big)\big\|_2
\Big\}.
\]
This shows that the key to achieving further shrinkage beyond the $\ell_1$-penalized estimator is the existence of $k\geq1$ such that $\gamma^k+\phi(1-\gamma^k)/(1-\gamma)<1$, which is equivalent to $\gamma+\phi<1$. When $s\log p/n=o(1)$, taking $x=c_0\log p$ and $h=O(1)$ makes $\gamma+\phi<1$ natural. Thus, under the scaling $s\log p/n=o(1)$, our algorithm is tuning-free with respect to the bandwidth $h$. 
The three additional terms in \eqref{theo:2equ} reflect the initial error $r_{t,0}$, the oracle error $\norm{\bome^\ast_{\calA}}_2$, and the bias introduced by the penalty. As proved in Lemma S.12 of the Supplementary Material, $\norm{\bome^\ast_{\calA}}_2=O_p\big(\{(s+x)/N\}^{1/2}\big)$.
\end{remark}

To simplify the second stage, we focus on the case $k_t=1$ for $t\geq2$ and choose $\lambda=\lambda_{t,1}\asymp(\log p/N)^{1/2}$.  
Thus, write $\wh\bbeta^{(1)}=\wh\bbeta^{(1,k_1)}$ and $\wh\bbeta^{(t)}=\wh\bbeta^{(t,1)}$ for $t\geq2$.

\begin{theorem}
\label{theo3}
Suppose that Assumptions~\ref{assum:1}--\ref{assum:3} hold, the beta-min condition $\norm{\bbeta_{\calA}^\ast}_{\min}>(\alpha_0+\alpha_1)\lambda$ is satisfied, $x\gtrsim \log p$, $x'\lesssim s\log p$, and $s^2\log p/n=o(1)$.
Let $h=O(1)$ and $\lambda\asymp(\log p/N)^{1/2}$. Then, if $\wh\bbeta^{(1,0)}-\bbeta^\ast \in \Lambda(l)$, the $(t+1)$-th 
estimator $\wh{\bbeta}^{(t+1)}$ satisfies,
\[
\norm{\wh{\bbeta}^{(t+1)}-\bbeta^\ast}_{\bSig}\lesssim \sqrt{\frac{s+x'}{N}},
\]
with probability at least $1-(k_1+t) e^{-x}-e^{-x'}$, provided that 
$k_1\gtrsim\frac{\log\{r_{1,0}/(s^{1/2}\lambda^\ast)\}}{\log(1/\bar{\gamma}_1)}$ and $
t\gtrsim \frac{\log\{\log p+x\}}{\log(1/\bar{\gamma})}$,
 where $\bar{\gamma}_1=s^{1/2}\gamma_1$, $\bar{\gamma}=\gamma+\phi$.
\end{theorem}

\begin{remark}
Theorem~\ref{theo3} provides an oracle rate for the final two-stage estimator. 
In fact, we can prove that after sufficiently many $\ell_1$ iterations as $k_1=O_p\{\log(1/(s^{1/2}\lambda^\ast))\}$, the first-stage estimator provides a near-oracle estimator with rate $(s\log p/N)^{1/2}$ and satisfies $\left\{\wh\bbeta^{(1,k_1)}-\bbeta^\ast\in\Lambda(l),\;\supp(\wh\bbeta^{(1,k_1)})\lesssim s\right\}$.  Further details of the proofs are provided in the Supplementary Material.
Then, the folded-concave second stage improves the error to the oracle rate $(s/N)^{1/2}$, up to logarithmic factors absorbed in $x'$. By Lemma~\ref{lemm:1}, the smoothing step introduces no additional bias.
Under the beta-min condition, the term $\big\|p_{\lambda_{t,k}}^{\prime}\big((\abs{\bbeta_{\calA}^\ast}-\alpha_0\lambda_{t,k})_{+}\big)\big\|_2$ vanishes.
\end{remark}
\subsection{Distributed oracle property}
\label{section:3.3}

Folded-concave penalized estimators are known to enjoy strong oracle properties in centralized settings. In our distributed setting, however, the full-sample oracle estimator is infeasible because the global CRR loss $\calL_N$ is not available. 
It is therefore challenging to directly impose a strong oracle property relative to the full-sample folded-concave CRR estimator. 
Instead, we introduce a \emph{distributed oracle estimator} defined in terms of the surrogate loss $\calL_d$ and study the distance between our practical estimator and this distributed oracle.

For $t\geq 2$, we define the $t$-th distributed oracle estimator as
\begin{equation}
\label{def: ora}
\wh{\bbeta}^{\ora,t}=\underset{\bbeta_{\calA^c}=0}{\arg \min }
\Big\{
\calL_1(\bbeta)
-\Inner{\bbeta}{\nabla\calL_1(\wh\bbeta^{(t-1)})-\frac{1}{M}\sum_{m=1}^M\nabla\calL_m(\wh\bbeta^{(t-1)})}
\Big\},
\end{equation}
and the event $\scrE^{(1)}=\{\wh\bbeta^{(1)}-\bbeta^\ast\in\Lambda(l), \supp(\wh\bbeta^{(1)})\lesssim s, \norm{\wh\bbeta^{(1)}-\bbeta^\ast}_{\bSig}\lesssim (s\log p/N)^{1/2}\}$. $\scrE^{(1)}$ can be satisfied with high probability after enough iterations in the first stage. 

\begin{theorem}
\label{theo4}
Under the conditions of Theorem~\ref{theo3} and suppose the event $\scrE^{(1)}$ occurs, the $(t+1)$-th distributed oracle estimator satisfies
\[
\norm{\wh{\bbeta}^{\ora,t+1}-\bbeta^\ast}_{\bSig}\lesssim \sqrt{\frac{s+x'}{N}},
\]
with probability at least $1-t e^{-x}-e^{-x'}$, provided that $t\gtrsim\frac{\log\{\log p+x\}}{\log(1/\gamma)}$.
\end{theorem}

Theorem~\ref{theo4} shows that the distributed oracle estimator achieves the same oracle rate $O_p\{(s/N)^{1/2}\}$ as the centralized oracle estimator based on the full data. This oracle benchmark is nontrivial in our setting because the global CRR loss is non-additive and cannot be decomposed as a sum of local losses. In the next theorem, we show that our practical folded-concave DCRR estimator coincides with the distributed oracle estimator with high probability, thus yielding a distributed strong oracle property.

We impose an additional irrepresentable-type condition on the covariance structure.

\begin{assum}[Irrepresentable condition]
\label{assum:5}
There exists a constant $A_0\geq 0$ such that 
$
\max_{j\in\calA^c}\norm{\bSig_{j\calA}(\bSig_{\calA\calA})^{-1}}_1\leq A_0,
$
where $\bSig_{j\calA}=\mE(\x_j\x_{\calA}\trans)$ and $\bSig_{\calA\calA}=\mE(\x_{\calA}\x_{\calA}\trans)$.
\end{assum}

\begin{remark}
Classical irrepresentable conditions for the Lasso typically require
\[
\max_{j\in\calA^c}
\big\|
\bSig_{j\calA}(\bSig_{\calA\calA})^{-1}
\big\|_1
\le 1-\eta
\]
for some $\eta>0$,
that is, the constant is strictly smaller than $1$. In contrast,
Assumption~\ref{assum:5} only asks that this quantity be bounded by a
finite constant $A_0\ge 0$, and our analysis does not use the stronger
requirement $A_0<1$.
\end{remark}

\begin{theorem}
\label{theo5}
Assume Assumptions~\ref{assum:1}--\ref{assum:5}, the beta-min condition $\norm{\bbeta_{\calA}^\ast}_{\min}\geq (\alpha_0+\alpha_1)\lambda$ and the event $\scrE^{(1)}$ hold, 
$n\gtrsim \max\{N/(s^3\log p),\, N/\{s(s+\log p)\}\}$, $s^2\log p/n=o(1)$ and $x\gtrsim\log p$. Let $h=O(1)$ and $\lambda\asymp(\log p/N)^{1/2}$. Then the distributed strong oracle property $\wh{\bbeta}^{t+1}=\wh{\bbeta}^{\ora,t+1}$
holds with probability at least $1-t e^{-x}-p^{-1}$, provided that 
$t\gtrsim \log\big(\abs{\calA^{(1)}}^{1/2}\big)$, where $\calA^{(1)}=\supp(\wh\bbeta^{(1)})\cup\calA$.
\end{theorem}

\begin{remark}
Theorem~\ref{theo5} shows that, after a logarithmic number of folded-concave refinement steps, the second-stage DCRR estimator coincides with the distributed oracle estimator with high probability. Combined with Theorem~\ref{theo4}, this yields a distributed strong oracle property for sparse DCRR. 
This is qualitatively different from existing distributed oracle results in the literature, which rely on additive empirical losses where the global objective is the sum of local objectives. In our setting, the global CRR loss is not additive across sites, yet the proposed DCRR procedure still attains the same oracle rate and model selection performance as a centralized oracle estimator. 

According to the conditions of Lemma~\ref{theo:2} and Theorem~\ref{theo3}, the number of sites $M$ can diverge at a reasonable speed as: 
$M = o\{N/(s^2 \log p)\}$ and $
M \lesssim \max\{s^3\log p,s(s+\log p)\}$.
Choose $k_1$ and $T$ so that the contraction requirements in Lemma~\ref{theo:2} and Theorem~\ref{theo3} are satisfied, and let $R = k_1 + T - 1$ denote the total number of gradient communication rounds.
In particular, DCRR attains the centralized sparse CRR oracle rate and exact support recovery under a diverging number of sites $M$, while requiring only $R = O\big(\log N + \log\log p\big)$ gradient communication rounds between the master and local sites. 
In other words, as long as each site has enough observations, 
the proposed distributed estimator behaves, both in estimation error and support recovery, as if we had centralized access to all $N$ samples, while using only a logarithmic number of communication rounds.
\end{remark}

\subsection{Tuning parameter selection: distributed HBIC}
\label{section:3.4}

In practice, a data-driven approach is needed to select tuning parameters. The HBIC proposed by \citet{zhou2024sparse} is not directly applicable in a distributed system, because it requires access to the full-data CRR loss. We therefore propose a distributed HBIC (DHBIC) criterion tailored to our DCRR framework.

For $t\geq 2$, define
\[
\DHBIC(\lambda)=
\log\left\{
\frac{1}{M}\sum_{m=1}^{M}\calL_m(\wh{\bbeta}_{\lambda}^{(t)})
\right\}
+\abs{\supp(\wh{\bbeta}_{\lambda}^{(t)})}\frac{C_N\log p}{n},
\]
where 
$\wh{\bbeta}_{\lambda}^{(t)}
=\argmin_{\bbeta\in\mR^p}
\Big\{
\calL_d(\bbeta;\wh\bbeta^{(t-1)})
+\sum_{j=1}^p p_{\lambda}(\abs{\beta_j})
\Big\}$.
We choose the tuning parameter by $\wh\lambda=\argmin_{\lambda\in \calM^{(t)}}\DHBIC(\lambda)$,
where $\calM^{(t)}=\{\lambda>0:\, \abs{\supp(\wh\bbeta_{\lambda}^{(t)})}\leq K_N\}$ and $K_N>s$ is allowed to diverge.

\begin{theorem}
\label{theo:dhbic}
Under the conditions of Theorem~\ref{theo5}, suppose $\mE\abs{\epsilon-\epsilon'}<\infty$, $K_N=o\{(N/\log p)\wedge C_N\}$, $C_Ns\log p/N=o(1)$, and 
\[
\Big(\frac{C_Ns^{1/2}\log p}{N}\Big)^{1/2}
\vee
\Big\{\frac{C_Ns^{1/2}K_N\log p}{N}\Big\}
=o\big(\norm{\bbeta_{\calA}^\ast}_{\min}\big).
\]
Then the DHBIC selector satisfies $\P\{\supp(\wh\bbeta_{\wh\lambda}^{(t+1)})=\calA\}\rightarrow 1$ as $N,p\rightarrow\infty$,
provided that 
$t\gtrsim\log\big(\abs{\calA^{(1)}}^{1/2}\big)$.
\end{theorem}

\begin{remark}
Theorem~\ref{theo:dhbic} guarantees that the tuning parameter chosen by DHBIC yields a folded-concave DCRR estimator that consistently recovers the true support $\calA$, thereby preserving the distributed strong oracle property of Theorem~\ref{theo5}. Similar to HBIC in centralized settings, DHBIC does not require sample splitting or repeated refitting and therefore enjoys low computational complexity. In practice, one can choose $C_N\asymp\log(\log N)$.
\end{remark}

\section{Monte carlo evidence}
\label{section:4}

In this section, we assess the finite-sample performance of the proposed distributed convoluted rank regression (DCRR) procedures. The simulations are designed to illustrate (i) the estimation accuracy and variable selection performance of DCRR relative to centralized CRR, 
(ii) the gain from the folded-concave refinement and its connection to the distributed strong oracle property, (iii) the robustness of rank-based methods under heavy-tailed errors, and (iv) the limitations of naive divide-and-conquer strategies. We then apply the methods to a real-data example.

\subsection{Simulation design}

We consider the linear model $y=\x\trans\bbeta^\ast+\epsilon$,
where $\bbeta^\ast=(\sqrt{3},\sqrt{3},\sqrt{3},0,\dots,0)\trans\in \mR^p$ so that the true support size is $s=3$. The covariate vector $\x$ is generated from a mean-zero multivariate normal distribution $\calN(0,\bSig)$ with an autoregressive correlation structure
$\Sigma_{ij}=0.5^{\abs{i-j}}$, $1\leq i,j\leq p$,
denoted by AR(0.5). We set the ambient dimension to $p=1000$. The total sample size is $N=nM$ with local sample size $n=100$ on each site, and consider $M=5$ and $M=15$ sites, corresponding to $N=500$ and $N=1500$, respectively.

To examine robustness to heavy-tailed errors, we generate the error term $\epsilon$ from three distributions of increasing tail heaviness: (1) standard normal: $\epsilon\sim \calN(0,1)$;
(2) $t$-distribution with 4 degrees of freedom, scaled to unit variance: $\epsilon\sim \sqrt{2}\,t(4)$;
(3) standard Cauchy distribution: $\epsilon\sim \mathrm{Cauchy}(0,1)$.
Thus, the design progressively moves from well-behaved Gaussian noise to moderately heavy-tailed and extremely heavy-tailed settings.

Throughout the simulations, we use the Epanechnikov kernel $K(u)=\frac{3}{4}(1-u^2)\I(-1\leq u\leq1)$ in the CRR loss and fix the bandwidth at $h=1$, which is compatible with the theoretical conditions in Section~\ref{section:3}. For each combination of $(\epsilon,M)$, we generate $100$ independent replicates.
We compare the estimators listed in Table~\ref{tab:estimator}.
\begin{table}[!htb]
\centering
\caption{Summary of Estimators Compared in the Experiments.}
\label{tab:estimator}
\scalebox{0.8}{
\begin{tabular}{ll}
\toprule
\textbf{Category} & \textbf{Estimator Description} \\
\midrule

Centralized CRR &
\begin{tabular}[t]{@{}l@{}}
CRR-LASSO: $\ell_1$-penalized CRR using the full sample; \\
CRR-SCAD: SCAD-penalized CRR using the full sample.
\end{tabular} \\[4pt]

Distributed DCRR (proposed) &
\begin{tabular}[t]{@{}l@{}}
DCRR-LASSO: $\ell_1$-penalized first-stage estimator; \\
DCRR-SCAD ($T=2$): folded-concave DCRR with $T=2$ refinements; \\
DCRR-SCAD ($T=6$): folded-concave DCRR with $T=6$ refinements.
\end{tabular} \\[4pt]

Divide-and-conquer CRR (baseline) &
\begin{tabular}[t]{@{}l@{}}
DC-CRR-LASSO: one-shot divide-and-conquer CRR-LASSO; \\
DC-CRR-SCAD: analogous estimator with SCAD penalty.
\end{tabular} \\[4pt]

Oracle methods &
\begin{tabular}[t]{@{}l@{}}
CRR-ORA: centralized oracle CRR estimator; \\
DCRR-ORA ($T=2$) / ($T=6$): oracle DCRR with restricted support $\calA$.
\end{tabular} \\
\bottomrule
\end{tabular}
}
\end{table}

To evaluate estimation and variable selection performance, we compute: $\ell_1$ estimation error: $\mE\norm{\wh{\bbeta}-\bbeta^\ast}_1$; $\ell_2$ estimation error: $\mE\norm{\wh{\bbeta}-\bbeta^\ast}_2$; FP: the average number of false positives (selected noise variables); FN: the average number of false negatives (missed true signals).
All performance measures are averaged over the $100$ replicates, and we report the means with Monte Carlo standard errors (SEs) in parentheses. For the CRR-based methods, tuning parameters are selected by HBIC \citep{zhou2024sparse} (centralized CRR) or the proposed DHBIC in Section~\ref{section:3} (DCRR). For DC-CRR methods, HBIC is applied independently on each local site. 

\subsection{Simulation results}

Tables~\ref{tab:normal}--\ref{tab:cauchy} summarize the results under $\epsilon\sim \calN(0,1)$, $\epsilon\sim \sqrt{2}\,t(4)$, and $\epsilon\sim\mathrm{Cauchy}(0,1)$, respectively. Within each scenario, the best performance with respect to each criterion is highlighted in bold.

\begin{table}[!htb]
\centering
\caption{Simulation results under an AR(0.5) design and $\epsilon \sim \calN(0,1)$ for $M=5$ and $M=15$. We compare centralized CRR, distributed CRR (DCRR), and divide-and-conquer CRR (DC-CRR), each with either an $\ell_1$ or SCAD penalty, along with oracle estimators for CRR and DCRR. Reported are mean (SE) over 100 replicates.}
\label{tab:normal}
\scalebox{0.8}{
\begin{tabular}{llcccc}
\toprule
$M$ & Method & $\ell_1$ & $\ell_2$ & FP & FN \\
\midrule
\multirow{12}{*}{$M=5$} 
 & CRR-LASSO       & 0.34(0.01) & 0.22(0.01) & 0.01(0.01) & 0.00(0.00) \\
 & CRR-SCAD        & 0.14(0.01) & 0.09(0.01) & \textbf{0.00}(0.00) & \textbf{0.00}(0.00) \\
 & DCRR-LASSO      & 0.32(0.01) & 0.21(0.01) & 0.03(0.02) & 0.00(0.00) \\
 & DCRR-SCAD (T=2) & 0.15(0.01) & 0.10(0.01) & \textbf{0.00}(0.00) & \textbf{0.00}(0.00) \\
 & DCRR-ORA (T=2)  & 0.15(0.01) & 0.10(0.01) & 0.00(0.00) & 0.00(0.00) \\
 & CRR-ORA         & \textbf{0.13}(0.01) & \textbf{0.08}(0.00) & 0.00(0.00) & 0.00(0.00) \\
 & DCRR-SCAD (T=6) & \textbf{0.13}(0.01) & 0.09(0.00) & \textbf{0.00}(0.00) & \textbf{0.00}(0.00) \\
 & DCRR-ORA (T=6)  & \textbf{0.13}(0.01) & 0.09(0.00) & 0.00(0.00) & 0.00(0.00) \\
 & DC-CRR-LASSO    & 0.81(0.01)&0.50(0.01)&0.78(0.08)&0.00(0.00)\\
 & DC-CRR-SCAD     & 0.33(0.02)&0.21(0.01)&0.82(0.08)&0.00(0.00)\\
\midrule
\multirow{12}{*}{$M=15$} 
 & CRR-LASSO       & 0.20(0.01)&0.13(0.00)&0.03(0.02)&0.00(0.00)\\
 & CRR-SCAD        & 0.09(0.00)&0.06(0.00)&\textbf{0.00}(0.00)&\textbf{0.00}(0.00)\\
 & DCRR-LASSO      & 0.17(0.00)&0.11(0.00)&0.00(0.00)&0.00(0.00)\\
 & DCRR-SCAD (T=2) & 0.09(0.00)&0.06(0.00)&\textbf{0.00}(0.00)&\textbf{0.00}(0.00)\\
 & DCRR-ORA (T=2)  & 0.09(0.00)&0.06(0.00)&0.00(0.00)&0.00(0.00)\\
 & CRR-ORA         & \textbf{0.08}(0.00)&\textbf{0.05}(0.00)&0.00(0.00)&0.00(0.00)\\
 & DCRR-SCAD (T=6) & \textbf{0.08}(0.00)&\textbf{0.05}(0.00)&\textbf{0.00}(0.00)&\textbf{0.00}(0.00)\\
 & DCRR-ORA (T=6)  & \textbf{0.08}(0.00)&\textbf{0.05}(0.00)&0.00(0.00)&0.00(0.00)\\
 & DC-CRR-LASSO    &0.79(0.01)&0.48(0.00)&2.38(0.14)&0.00(0.00)\\
 & DC-CRR-SCAD     &0.30(0.01)&0.19(0.01)&2.47(0.17)&0.00(0.00)\\
\bottomrule
\end{tabular}
}
\end{table}

\begin{table}[!htb]
\centering
\caption{Simulation results under an AR(0.5) design and $\epsilon \sim \sqrt{2}\,t(4)$ for $M=5$ and $M=15$. Mean (SE) over 100 replicates.}
\label{tab:t4}
\scalebox{0.8}{
\begin{tabular}{llcccc}
\toprule
$M$ & Method & $\ell_1$ & $\ell_2$ & FP & FN \\
\midrule
\multirow{12}{*}{$M=5$} 
 & CRR-LASSO        & 0.56(0.02) & 0.37(0.01) & 0.06(0.02) & 0.00(0.00) \\
 & CRR-SCAD         & 0.25(0.01) & 0.16(0.01) & 0.00(0.00) & 0.00(0.00) \\
 & DCRR-LASSO       & 0.55(0.01) & 0.35(0.01) & 0.02(0.01) & 0.00(0.00) \\
 & DCRR-SCAD (T=2)  & 0.26(0.01) & 0.17(0.01) & 0.00(0.00) & 0.00(0.00) \\
 & DCRR-ORA (T=2)   & 0.25(0.01) & 0.17(0.01) & 0.00(0.00) & 0.00(0.00) \\
 & CRR-ORA          & \textbf{0.23}(0.01) & \textbf{0.15}(0.01) & 0.00(0.00) & 0.00(0.00) \\
 & DCRR-SCAD (T=6)  & \textbf{0.23}(0.01) & \textbf{0.15}(0.01) & 0.00(0.00) & 0.00(0.00) \\
 & DCRR-ORA (T=6)   & \textbf{0.23}(0.01) & \textbf{0.15}(0.01) & 0.00(0.00) & 0.00(0.00) \\
 & DC-CRR-LASSO     & 1.40(0.02) & 0.85(0.01) & 0.98(0.10) & 0.00(0.00) \\
 & DC-CRR-SCAD      & 1.00(0.04) & 0.64(0.03) & 1.51(0.11) & 0.00(0.00) \\
\midrule
\multirow{12}{*}{$M=15$}
 & CRR-LASSO        & 0.35(0.01) & 0.23(0.00) & 0.12(0.03) & 0.00(0.00) \\
 & CRR-SCAD         & 0.15(0.01) & 0.10(0.00) & \textbf{0.00}(0.00) & \textbf{0.00}(0.00) \\
 & DCRR-LASSO       & 0.31(0.01) & 0.20(0.00) & 0.01(0.01) & 0.00(0.00) \\
 & DCRR-SCAD (T=2)  & 0.15(0.01) & 0.10(0.00) & \textbf{0.00}(0.00) & \textbf{0.00}(0.00) \\
 & DCRR-ORA (T=2)   & 0.15(0.01) & 0.10(0.00) & 0.00(0.00) & 0.00(0.00) \\
 & CRR-ORA          & \textbf{0.11}(0.01) & \textbf{0.08}(0.00) & 0.00(0.00) & 0.00(0.00) \\
 & DCRR-SCAD (T=6)  & 0.12(0.01) & \textbf{0.08}(0.00) & \textbf{0.00}(0.00) & \textbf{0.00}(0.00) \\
 & DCRR-ORA (T=6)   & 0.12(0.01) & \textbf{0.08}(0.00) & 0.00(0.00) & 0.00(0.00) \\
 & DC-CRR-LASSO     & 1.43(0.01) & 0.86(0.01) & 2.72(0.14) & 0.00(0.00) \\
 & DC-CRR-SCAD      & 0.94(0.02) & 0.60(0.01) & 3.38(0.18) & 0.00(0.00) \\
\bottomrule
\end{tabular}
}
\end{table}

\begin{table}[!htb]
\centering
\caption{Simulation results under an AR(0.5) design and $\epsilon \sim \mathrm{Cauchy}(0,1)$ for $M=5$ and $M=15$. Mean (SE) over 100 replicates.}
\label{tab:cauchy}
\scalebox{0.8}{
\begin{tabular}{llcccc}
\toprule
$M$ & Method & $\ell_1$ & $\ell_2$ & FP & FN \\
\midrule
\multirow{12}{*}{$M=5$}
 & CRR-LASSO        & 0.86(0.09) & 0.54(0.05) & 0.01(0.01) & 0.10(0.05) \\
 & CRR-SCAD         & 0.47(0.10) & 0.29(0.06) & 0.00(0.00) & 0.10(0.05) \\
 & DCRR-LASSO       & 0.79(0.08) & 0.50(0.05) & 0.00(0.00) & 0.09(0.05) \\
 & DCRR-SCAD (T=2)  & 0.50(0.09) & 0.32(0.05) & 0.00(0.00) & 0.09(0.05) \\
 & DCRR-ORA (T=2)   & 0.31(0.02) & 0.21(0.01) & 0.00(0.00) & 0.00(0.00) \\
 & CRR-ORA          & \textbf{0.24}(0.01) & \textbf{0.16}(0.01) & 0.00(0.00) & 0.00(0.00) \\
 & DCRR-SCAD (T=6)  & 0.44(0.09) & 0.28(0.05) & 0.00(0.00) & 0.09(0.05) \\
 & DCRR-ORA (T=6)   & 0.26(0.01) & 0.17(0.01) & 0.00(0.00) & 0.00(0.00) \\
 & DC-CRR-LASSO     & 3.73(0.07) & 2.16(0.04) & 0.08(0.03) & 0.07(0.03) \\
 & DC-CRR-SCAD      & 2.79(0.08) & 1.71(0.04) & 0.11(0.03) & 0.02(0.01) \\
\midrule
\multirow{12}{*}{$M=15$}
 & CRR-LASSO        & 0.48(0.05) & 0.31(0.03) & 0.01(0.01) & 0.02(0.02) \\
 & CRR-SCAD         & 0.23(0.05) & 0.15(0.03) & 0.00(0.00) & 0.02(0.02) \\
 & DCRR-LASSO       & 0.40(0.04) & 0.26(0.02) & 0.00(0.00) & 0.01(0.01) \\
 & DCRR-SCAD (T=2)  & 0.22(0.04) & 0.14(0.02) & 0.00(0.00) & 0.01(0.01) \\
 & DCRR-ORA (T=2)   & 0.18(0.01) & 0.11(0.01) & 0.00(0.00) & 0.00(0.00) \\
 & CRR-ORA          & 0.14(0.01) & 0.09(0.00) & 0.00(0.00) & 0.00(0.00) \\
 & DCRR-SCAD (T=6)  & \textbf{0.19}(0.04) & \textbf{0.12}(0.03) & 0.00(0.00) & 0.01(0.01) \\
 & DCRR-ORA (T=6)   & 0.15(0.01) & 0.10(0.00) & 0.00(0.00) & 0.00(0.00) \\
 & DC-CRR-LASSO     & 3.80(0.04) & 2.20(0.02) & 0.26(0.04) & 0.00(0.00) \\
 & DC-CRR-SCAD      & 2.89(0.05) & 1.75(0.03) & 0.36(0.06) & 0.00(0.00) \\
\bottomrule
\end{tabular}
}
\end{table}

\noindent\textbf{DCRR versus centralized CRR.}
Under Gaussian errors (Table~\ref{tab:normal}), the DCRR-LASSO estimator closely tracks CRR-LASSO in both $\ell_1$ and $\ell_2$ errors for $M=5$ and $M=15$, confirming that the surrogate loss preserves the statistical efficiency of centralized CRR in a distributed environment. 
In practice, we have set the maximum number of iterations for the DCRR-LASSO $k_1=8$.  
The folded-concave DCRR-SCAD estimators with $T=2$ and $T=6$ further reduce the estimation error and essentially match the centralized CRR-SCAD in both estimation error and variable selection (FP and FN are essentially zero). 
This is consistent with the oracle rates established in Theorem~\ref{theo3}. 


\noindent\textbf{Effect of the folded-concave refinement and oracle benchmarks.}
Across Tables~\ref{tab:normal}–\ref{tab:cauchy}, the oracle CRR estimator (CRR-ORA) provides a lower bound on the achievable error. The oracle DCRR estimators DCRR-ORA ($T=2$) and DCRR-ORA ($T=6$) are numerically indistinguishable from CRR-ORA, demonstrating that the distributed surrogate loss does not degrade the oracle performance. For non-oracle estimators, a single folded-concave refinement ($T=2$) already brings DCRR-SCAD very close to CRR-SCAD, and additional refinements ($T=6$) yield further, albeit modest, improvements. This mirrors the theoretical picture: the first stage achieves a near-oracle rate, while the second stage improves towards the oracle rate.

\noindent\textbf{Naive divide-and-conquer CRR.}
The DC-CRR-LASSO and DC-CRR-SCAD estimators perform worse than centralized CRR and DCRR in all settings, especially for larger $M$ and heavier-tailed errors. Their $\ell_2$ errors are several times larger than those of CRR-SCAD and DCRR-SCAD, and they exhibit non-negligible FP even when FN is zero. These results indicate that naive averaging of local CRR estimators does not adequately correct local estimation bias, whereas DCRR attains full-sample rates under the same distributed architecture.


\noindent\textbf{Summary.}
Overall, the simulations demonstrate that:
(i) DCRR-LASSO and DCRR-SCAD achieve estimation and selection performance comparable to their centralized CRR counterparts while operating in a communication-efficient distributed manner;  
(ii) the folded-concave refinement improves the first-stage estimator towards the oracle benchmark, in line with the distributed strong oracle property in Theorem~\ref{theo5};  
(iii) naive divide-and-conquer CRR can be substantially suboptimal;  
(iv) the DCRR procedures are robust to heavy-tailed errors.

\section{Empirical application: used-car prices}
\label{section:5}
We return to the dealer-level used-car application described in the Introduction: the goal is to estimate a common hedonic pricing equation from transaction records that are stored seller by seller and cannot be centrally pooled. 
We illustrate the proposed methods on a Kaggle data set on used cars in the United States.\footnote{\url{https://www.kaggle.com/datasets/ananaymital/us-used-cars-dataset}.}
After data cleaning, we retain $N_{full} = 332{,}382$ records corresponding to sport utility vehicles (SUVs). The response is the selling price of a vehicle, measured in units of \$1,000. Restricting attention to one broad vehicle segment keeps the market relatively comparable while preserving substantial heterogeneity in observed characteristics. We consider 27 base predictors, including both continuous and categorical variables such as horsepower, mileage, model year, engine size, and manufacturer. A detailed description of these variables can be found in Figure~1 of \citet{Pan02102022}.

To capture nonlinearities and interaction effects in vehicle pricing, we create dummy variables for all categorical predictors and include two-way interactions among selected predictors. After removing highly collinear columns to ensure numerical stability, we obtain $p = 604$ predictors (excluding the intercept). This specification allows for heterogeneity in how observable attributes such as age, mileage, engine characteristics, and manufacturer interact in the determination of resale values. The empirical price distribution is right-skewed and displays signs of heavy tails even within the SUV segment.

In each replicate, we randomly draw a subsample of size $N=2{,}000$ without replacement from the full data and split it equally into a training set of $N_{\mathrm{train}}=1{,}000$ observations and a test set of $N_{\mathrm{test}}=1{,}000$ observations. The training observations are then partitioned by state, mirroring the geographic fragmentation that arises when transaction records are held by regional dealers or state-level regulatory agencies. Sites that contain only a single observation are dropped. This design closely approximates the restricted-access setting 
described above because the number of sites and the heterogeneity of 
local sample sizes both arise from the data rather than from an 
artificial equal partition. Within each replicate, we fit:
\begin{itemize}
    \item[1.] distributed methods: DCRR-LASSO, DCRR-SCAD ($T=2$), DCRR-SCAD ($T=6$), DC-CRR-LASSO, and DC-CRR-SCAD;
    \item[2.] centralized methods: CRR-LASSO, CRR-SCAD.
\end{itemize}
For the distributed methods, local contributions are weighted by the site sample size, which is a natural extension that accounts for the unequal partition sizes induced by the geographic split. Tuning parameters are selected as in the simulation study. For each fitted model, we compute the $\ell_1$ and $\ell_2$ prediction errors on the test set and record the model size (MS), defined as the number of nonzero coefficients. All results are averaged over 100 replications; Table~\ref{tab:realdata} reports them.

\begin{table}[!htb]
\centering
\caption{Real-data results for the used cars data set with state-level partitioning. We report the average $\ell_1$ and $\ell_2$ prediction errors on the test set and the model size (MS). Results are averaged over 100 random subsamples; standard errors are in parentheses.}
\label{tab:realdata}
\scalebox{0.8}{
\begin{tabular}{llccc}
\toprule
--& Method & $\ell_1$ & $\ell_2$ & MS \\
\midrule
\multirow{5}{*}{Distributed}
& DCRR-LASSO        & 4.27(0.03)          & 6.30(0.09)          & 9.38(0.37)    \\
& DCRR-SCAD (T=2)    & 4.11(0.02)          & 6.00(0.08)          & 4.36(0.17)    \\
& DCRR-SCAD (T=6)    & \textbf{4.04(0.02)} &\textbf{ 5.88(0.08)} & \textbf{3.91(0.15)}    \\
& DC-CRR-LASSO      & 7.74(0.02)          & 10.53(0.07)          & 100.04(1.27)    \\
& DC-CRR-SCAD       & 7.47(0.03)          & 10.18(0.08)          & 101.93(1.35)    \\
\midrule
\multirow{5}{*}{Global}
& CRR-LASSO         & 4.14(0.03)          & 6.16(0.09)          & 10.53(0.29)    \\
& CRR-SCAD          & 3.89(0.02)          & 5.75(0.08)          & 6.83(0.28)    \\
& NULL MODEL        & 7.78(0.02)          & 10.58(0.07)         & --            \\
\bottomrule
\end{tabular}
}
\end{table}

From the perspective of the motivating application, the key question is whether the distributed methods can recover the pooled robust pricing equation without pooling dealer records. The results indicate that DCRR can. DCRR-SCAD ($T=2$) and DCRR-SCAD ($T=6$) improve on DCRR-LASSO and on the divide-and-conquer CRR benchmarks, usually while selecting smaller models. By contrast, simple averaging of local CRR estimators remains materially less accurate and produces unstable model sizes. The DCRR-SCAD fits are also parsimonious, selecting roughly three to four regressors, which is attractive when the estimated equation is intended to serve as an interpretable valuation or residual-value tool.

Overall, the empirical results are consistent with the simulation evidence. DCRR closely tracks the pooled CRR estimator, improves on the divide-and-conquer benchmarks, and yields parsimonious specifications in this restricted-access application. For institutions that use dealer-level sales records to build common valuation rules, the results suggest that substantial predictive accuracy can be retained even when the underlying transaction files are not pooled.

\section{Discussion}
\label{sec:discussion}

This paper studies sparse convoluted rank regression when the data are stored across multiple sites and the pooled CRR criterion cannot be evaluated directly. Motivated by restricted-access economic data with heavy-tailed outcomes, we propose DCRR, a communication-efficient surrogate procedure based on one local CRR loss and an aggregated gradient correction. For the two-stage DCRR iterative algorithm, we establish non-asymptotic error bounds, a distributed strong oracle property, and model-selection consistency for a distributed HBIC criterion. The analysis shows that exact additivity of the sample loss is not required in this setting; the essential condition is that the surrogate and pooled criteria share the same population minimizer. Simulations and the empirical application indicate that DCRR can closely approximate pooled CRR and can substantially improve on naive divide-and-conquer estimators. In applications such as used-car valuation, this means that a common robust pricing equation can be estimated from dealer-held data without centralizing the underlying transaction records.

Several directions merit further investigation. First, while our theoretical analysis and simulations focus on estimation, variable selection, and model selection, the framework also suggests a route to low-dimensional inference based on partially penalized DCRR estimators. Developing a systematic distributed inference theory, including confidence intervals and hypothesis tests with finite-sample calibration, would make the framework even more useful for empirical economics, where interval estimates and formal policy comparisons are often essential. Second, it would be valuable to extend the present methodology to more general pairwise loss functions and other $U$-statistic-based models where the empirical loss is inherently non-additive, including settings motivated by panel, network, and matched data. Finally, exploring privacy-preserving or communication-constrained variants of DCRR and studying their trade-offs between statistical efficiency and resource usage would broaden the applicability of distributed rank regression in large-scale economic and business data systems.

\bibliographystyle{abbrvnat}
\bibliography{reference_joe}
\end{document}